\documentstyle[aps,multicol,eqsecnum]{revtex}

\widetext
\renewcommand{\narrowtext}{\begin{multicols}{2} \global\columnwidth20.5pc}
\renewcommand{\widetext}{\end{multicols} \global\columnwidth42.5pc}

\begin{document}
\draft

\title{Thermodynamic potential of interacting Bose-Einstein gas \\
       confined in harmonic potential}

\author{Makoto Hiramoto}

\address{College of Science and Technology, Nihon University, 
         Funabashi, Chiba 274-8501, Japan}

\date{\today}

\maketitle

\begin{abstract}
We investigate the interaction effect between atoms and the finite size effect 
of a Bose-Einstein gas at finite temperature. Using a mean field approach, we 
derive the thermodynamic potential on finite systems and obtain the condensate 
fraction, the chemical potential and the internal energy with the interaction 
and the finite size effects taken into account. In this approach, we can treat 
thermodynamic quantities as those of the ideal Bose gas. It is found that the 
interaction effect shifts the transition temperature by a few percent toward 
the lower temperature. Above the transition temperature, these quantities agree 
with the noninteracting Bose gas.
\end{abstract}

\pacs{PACS numbers: 03.75.Fi, 05.30.Jp, 32.80.Pj, 51.30.+i}

\narrowtext
\section{Introduction}
\label{sec:level1}

Bose-Einstein condensation (BEC) was first observed in dilute ultracold alkali 
atoms of rubidium \cite{and}, lithium \cite{bra} and sodium \cite{dav}. 
Furthermore, Mewes et al. and Ensher et al. measured the condensate fraction 
and the energy of rubidium atoms  \cite{mew,ens}. It revealed that the 
transition temperature $T_{0}$ is shifted not only by the interaction effect 
but also by the finite size effect by a few percent, where $T_{0}$ denotes the 
transition temperature of the noninteracting Bose gas within the external field 
in the thermodynamic limit. This means that both the interaction and the finite 
size effects play an important role in the real Bose gases. In fact, the number 
of trapped atoms is typically $N = 10^3 \sim 10^6$, and this may not be 
sufficiently large to take the thermodynamic limit. For the noninteracting Bose 
gas, the condensate fraction confined in a spherical harmonic potential can be 
analytically given as \cite{ket,hau},
\begin{eqnarray}
\frac{N_0}{N} & = & 1-\biggl(\frac{T}{T_0}\biggl)^3-\frac{3}{2}
\frac{\zeta(2)}{\zeta(3)^{2/3}}\frac{1}{N^{1/3}}\biggl(\frac{T}{T_0}\biggl)^2,
\label{eq1}
\end{eqnarray}
where the first two terms in the right hand side correspond to the condensate 
fraction in the thermodynamic limit and the third term is the finite size 
correction.

On the other hand, these experimental findings have stimulated much interest 
in the theory of the interacting Bose gas. Since BEC occurs when atoms are 
dilute and cold, we can treat the interaction of atoms as the two-body 
interacting Bose gas. In this case, the s-wave scattering length characterizes 
the strength of the two-body interaction. 
Under these conditions, the Gross-Pitaevskii (GP) equation can well describe 
the behavior of the interacting Bose gas at zero temperature \cite{gro,pit}. 
This is a mean field approach for the order parameter associated with the 
condensate. Using the GP equation, several authors \cite{dal} studied the 
ground state and the excitation properties of the condensate.

To study BEC at finite temperature, the Bogoliubov approximation at zero 
temperature was extended by Griffin \cite{gri}. It is called the 
Hartree-Fock-Bogoliubov (HFB) theory. In particular, Popov approximation to 
the HFB theory has been employed to explain the experimental results 
\cite{gioa}. This approximation neglect the anomalous average of the 
fluctuating field operator in the HFB theory. With the Popov approximation, 
thermodynamic quantities, such as the condensate fraction and the internal 
energy, agree with the experimental results at low temperature. However, recent 
experimental data of the excitation frequencies of the condensates indicate a 
discrepancy with these approximate results near the transition temperature 
$T_{0}$  $(T/T_{0}>0.6)$ \cite{jin}.

In this paper, we derive the thermodynamic potential for the interacting Bose 
gas within a mean field approximation which is different from Bogoliubov 
approximation. With this thermodynamic potential, we obtain the thermodynamic 
quantities; the condensate fraction, the entropy and the internal energy. These 
quantities have the same form as that of the ideal Bose gas, but the energy 
$E_{\bf i}$ contains the interaction effect and varies as a function of the 
temperature $T$ and the total number $N$. Therefore, we must determine not only 
the chemical potential $\mu$ but also the energy $E_{\bf i}$ as a function of 
$T$ and $N$ in a self-consistent fashion. In particular, the chemical potential 
determines the transition temperature on the condition that the chemical 
potential $\mu$ is equal to the ground state energy $E_{0}$ at $T=T_{0}$. Here, 
we choose the repulsive interaction where the scattering length $a$ is 
positive, and numerically solve the chemical potential and the energy 
$E_{{\bf i}}$. By determining the chemical potential near $T \sim T_{0}$, we 
find that the interaction effect shifts $T_{0}$ by a few percent toward the 
lower temperature which should be compared with the previous estimations of the 
semiclassical approximation \cite{giob}.

This paper is organized in the following way. In the next section, we derive 
the thermodynamic potential for the interacting Bose gas at finite temperature. 
Then, we obtain thermodynamic quantities in a self-consistent way. In 
section 3, we carry out the numerical evaluation of the chemical potential, the 
condensate fraction and the internal energy, in particular near $T_{0}$. Section 4 summarizes what we have clarified in this paper.

\section{Thermodynamics for Interacting Bose Gas}
\label{sec:level2}

We derive the thermodynamic potential for the interacting Bose gas within 
a mean field approach. To begin with, we construct the effective Hamiltonian 
in this system. The Hamiltonian for the interacting Bose gas confined in an 
external potential $V_{\rm{ext}}$ can be written as
\begin{eqnarray}
\hat{H} & = & \int d{\bf r}\hat{\Psi}^{\dagger}({\bf r})
\left[-\frac{\hbar^2}{2m}\nabla^2+V_{\rm{ext}}({\bf r})\right]
\hat{\Psi}({\bf r})
\nonumber \\
& & +\frac{1}{2}\int d{\bf r}d{\bf r}'
\hat{\Psi}^{\dagger}({\bf r})\hat{\Psi}^{\dagger}({\bf r'})V({\bf r}-{\bf r'})
\hat{\Psi}({\bf r'})\hat{\Psi}({\bf r}),
\label{eq2}
\end{eqnarray}
where $\hat{\Psi}({\bf r})$ is the boson field operator and 
$V({\bf r}-{\bf r'})$ is the two-body atomic potential. Here, we choose the 
spherical harmonic potential for the $V_{\rm{ext}}$ for simplicity. Thus, the 
$V_{\rm{ext}}$ is given as
\begin{eqnarray}
V_{\rm{ext}}({\bf r}) & = & \frac{1}{2}m\omega^2 r^2.
\label{eq3}
\end{eqnarray}
Since we are interested in the dilute and cold Bose gas, we replace 
$V({\bf r}-{\bf r'})$ by an effective interaction of the $\delta$-function type,
\begin{eqnarray}
V({\bf r}-{\bf r'}) & = & g\delta^{(3)}({\bf r-r'}),
\label{eq4}
\end{eqnarray}
where the coupling constant $g$ is related to the s-wave scattering length $a$ 
through
\begin{eqnarray}
g & = & \frac{4\pi\hbar^2a}{m}.
\label{eq5}
\end{eqnarray}
The scattering length is an important length scale in this system. A typical 
value of the scattering length $a$ is of the order of $nm$. The boson field 
operator for the ideal system can be written as
\begin{eqnarray}
\hat{\Psi}({\bf r}) & = & \frac{1}{\sqrt{V}}\sum_{{\bf k}}a_{{\bf k}}
e^{i{\bf k}\cdot{\bf r}}.
\label{eq6}
\end{eqnarray}
But for the general case, the corresponding field operator is a sum over all 
normal modes \cite{fet}
\begin{eqnarray}
\hat{\Psi}({\bf r}) & = & \sum_{{\bf n}}a_{{\bf n}}\chi_{{\bf n}}({\bf r}),
\label{eq7}
\end{eqnarray}
when $\chi_{\bf n}(\bf r)$'s define any complete set of normalized 
single-particle wave function $a_{{\bf n}}$ is a bosonic annihilation operator 
for the single-particle state ${\bf n}$. In our case, $\chi_{{\bf n}}({\bf r})$ 
is given as
\begin{eqnarray}
\chi_{{\bf n}}({\bf r}) & \equiv & u_{n_{x}n_{y}n_{z}}(x,y,z) \nonumber \\
& = & N_{n_{x}}N_{n_{y}}N_{n_{z}}e^{-\frac{1}{2}(\frac{r}{a_{ho}})^2}
\nonumber \\ & & 
H_{n_{x}}(x/a_{ho})H_{n_{y}}(y/a_{ho})H_{n_{z}}(z/a_{ho}),
\label{eq8}
\end{eqnarray}
where $H_{n}(\xi)$ is the Hermite polynomial, and $N_{n}$ is the normalization 
constant which can be written as
\begin{eqnarray}
N_{n} & = & \sqrt{\frac{a_{ho}^{-1}}{\pi^{1/2}2^n n!}},
\label{eq9}
\end{eqnarray}
where $a_{ho}$ denotes the harmonic oscillator length 
\begin{eqnarray}
a_{ho} & = & \sqrt{\frac{\hbar}{m\omega}}.
\label{eq10}
\end{eqnarray}
This is also an important length scale in this system. It is typically of the 
order of $\mu m$. From Eqs.\ (\ref{eq2}) and \ (\ref{eq7}), 
the Hamiltonian can be written as
\begin{eqnarray}
\hat{H} & = & 
\sum_{{\bf n}} \varepsilon_{{\bf n}}a^{\dagger}_{{\bf n}}a_{{\bf n}}
\nonumber \\ & & 
+\frac{2\pi\hbar^2a}{m}\sum_{\{{\bf n_{i}}\}}
\langle {\bf n_{1}}{\bf n_{2}}\mid{\bf n_{3}}{\bf n_{4}}\rangle
a^{\dagger}_{{\bf n_{1}}}a^{\dagger}_{{\bf n_{2}}}
a_{{\bf n_{3}}}a_{{\bf n_{4}}},
\label{eq11}
\end{eqnarray}
with
\begin{eqnarray}
\varepsilon_{{\bf n}} & = & \hbar\omega
\left(n_{x}+n_{y}+n_{z}+\frac{3}{2}\right), \label{eq12}\\
\langle {\bf n_{1}}{\bf n_{2}}\mid{\bf n_{3}}{\bf n_{4}}\rangle 
& = & \int d{\bf r}\chi^{\dagger}_{{\bf n_{1}}}({\bf r})
\chi^{\dagger}_{{\bf n_{2}}}({\bf r})
\chi_{{\bf n_{3}}}({\bf r})\chi_{{\bf n_{4}}}({\bf r}).
\label{eq13}
\end{eqnarray}

Now we consider the first order perturbation theory for the thermodynamic 
potential \cite{lan}. The first order correction is given by the diagonal part 
of the interacting Hamiltonian in Eq.\ (\ref{eq11}). This corresponds to 
ignoring the anomalous average. Therefore, the effective Hamiltonian can be 
written as 
\begin{eqnarray}
\hat{H}_{eff} & = & \sum_{{\bf i}} \varepsilon_{{\bf i}}n_{{\bf i}}
+\hbar\omega\frac{4\pi a}{a_{ho}}\sum_{{\bf i,j}}U_{{\bf ij}}
n_{{\bf i}}n_{{\bf j}},
\label{eq14}
\end{eqnarray}
where $n_{{\bf i}}=a^{\dagger}_{{\bf i}}a_{{\bf i}}$ and 
\begin{eqnarray}
U_{{\bf ij}} & = & F_{i_{x},j_{x}}F_{i_{y},j_{y}}F_{i_{z},j_{z}}, \nonumber \\ 
F_{i,j} & = & 
\frac{1}{\pi 2^{i}2^{j}i!j!}\int d\xi e^{-2\xi^2}H^2_{i}(\xi)H^2_{j}(\xi).
\label{eq15}
\end{eqnarray}
Now we rewrite Eq.\ (\ref{eq14}) in the following way
\begin{eqnarray}
\hat{H}_{eff} & = & \hat{H}_{0}+\hat{H}',
\label{eq16}
\end{eqnarray}
with
\begin{eqnarray}
\hat{H}_{0} & = & \sum_{{\bf i}}E_{{\bf i}}n_{{\bf i}}-
\hbar\omega\frac{4\pi a}{a_{ho}}\sum_{{\bf i,j}}U_{{\bf ij}}
\nu_{\bf i}\nu_{\bf j},\label{eq17}\\
\hat{H}' & = & \hbar\omega\frac{4\pi a}{a_{ho}}\sum_{{\bf i,j}}U_{{\bf ij}}
(n_{{\bf i}}-\nu_{\bf i})
(n_{{\bf j}}-\nu_{\bf j}),\label{eq18}\\
E_{{\bf i}} & = & \varepsilon_{{\bf i}}+\hbar\omega\frac{8\pi a}{a_{ho}}
\sum_{{\bf j}}U_{{\bf ij}}\nu_{\bf j}.
\label{eq19}
\end{eqnarray}
Here, we minimize $\hat{H}'$ by taking $\nu_{\bf i}$ the mean occupation 
numbers
\begin{eqnarray}
\nu_{\bf i} = \langle n_{{\bf i}}\rangle &=& 
\frac{{\rm{Tr}}\ n_{{\bf i}}\ e^{-(\hat{H}_0-\mu \hat{N})/k_B T}}
{{\rm{Tr}}\ e^{-(\hat{H}_0-\mu \hat{N})/k_B T}} \nonumber \\&=& 
\frac{1}{e^{(E_{{\bf i}}-\mu )/k_B T}-1}. 
\label{eq20}
\end{eqnarray}
Then, the thermodynamic potential $\Omega=-pV$ can be given as
\begin{eqnarray}
\Omega & = & -k_{B}T\ln {\rm Tr}\ e^{-(\hat{H}_{0}-\mu \hat{N})/k_{B}T},
 \nonumber \\
& \approx & -\hbar\omega\frac{4\pi a}{a_{ho}}\sum_{{\bf i,j}}U_{{\bf ij}}
\langle n_{{\bf i}}\rangle\langle n_{{\bf j}}\rangle 
\nonumber \\ & &
+k_{B}T\sum_{{\bf i}}\ln{(1-e^{-(E_{{\bf i}}-\mu)/k_{B}T})},
\label{eq21}
\end{eqnarray}
where the first term in the right hand side is the first order correction. 
We note that Eq.\ (\ref{eq20}) can be also determined by the condition 
\begin{eqnarray}
0 & = & \left(\frac{\partial \Omega}{\partial \langle n_{{\bf i}}\rangle}\right)
_{T,\mu,V,\langle n_{{\bf j}}\rangle\ne \langle n_{{\bf i}}\rangle}
\nonumber \\ 
& = & -\sum_{{\bf j}}U_{{\bf ij}}\langle n_{{\bf j}}\rangle
+\sum_{{\bf j}}U_{{\bf ij}}\frac{1}{e^{(E_{{\bf j}}-\mu )/k_B T}-1}.
\label{eq22}
\end{eqnarray}
This condition means that the thermodynamic potential $\Omega$ is a minimum 
with respect to any change of state at constant $T$, $V$ and $\mu$ in a state 
of thermal equilibrium.

Now we derive the thermodynamic quantities from Eqs.\ (\ref{eq21}) and 
\ (\ref{eq22}). First, we 
obtain the total number $N$ from Eq.\ (\ref{eq21}) with the condition of 
Eq.\ (\ref{eq22})
\begin{eqnarray}
N & = & -\left(\frac{\partial \Omega}{\partial \mu}\right)_{T,V} \nonumber \\
& = & 
-\left(\frac{\partial \Omega}{\partial \mu}\right)_
{T,V,\mu,\langle n_{{\bf i}}\rangle}-\sum_{{\bf i}}
\left(\frac{\partial \Omega}{\partial \langle n_{{\bf i}}\rangle}\right)
\cdot
\left(\frac{\partial \langle n_{{\bf i}}\rangle}{\partial \mu }\right)
\nonumber \\
& = & \sum_{{\bf i}}\frac{1}{e^{(E_{{\bf i}}-\mu )/k_B T}-1}.
\label{eq23}
\end{eqnarray}
We can determine the chemical potential $\mu$ and the energy $E_{\bf i}$ from 
Eqs.\ (\ref{eq19}) and \ (\ref{eq23}) which should be solved self-consistently. Then, we obtain the thermodynamic quantities through Eq.\ (\ref{eq21}) with 
Eq.\ (\ref{eq22}). For example, the entropy $S$ can be given by
\begin{eqnarray}
S &=& -\left(\frac{\partial \Omega}{\partial T}\right)_{V,\mu} \nonumber \\ 
& = & 
-\left(\frac{\partial \Omega}{\partial T}\right)_
{V,\mu,\langle n_{{\bf i}}\rangle}-\sum_{{\bf i}}
\left(\frac{\partial \Omega}{\partial \langle n_{{\bf i}}\rangle}\right)
\cdot
\left(\frac{\partial \langle n_{{\bf i}}\rangle}{\partial T }\right)
\nonumber \\
&=& k_{B}\sum_{{\bf i}}\left[(1+\langle n_{{\bf i}}\rangle)
\ln (1+\langle n_{{\bf i}}\rangle)-
\langle n_{{\bf i}}\rangle\ln \langle n_{{\bf i}}\rangle\right],
\label{eq24}
\end{eqnarray}
where we also use the condition Eq.\ (\ref{eq22}). Finally, the internal 
energy $E$ can be given as 
\begin{eqnarray}
E &=& \sum_{{\bf i}}\frac{E_{{\bf i}}}{e^{(E_{{\bf i}}-\mu )/k_B T}-1}.
\label{eq25}
\end{eqnarray}

We obtain the thermodynamic quantities from Eq.\ (\ref{eq21}) by solving 
$E_{{\bf i}}$ and $\mu$ from Eqs.\ (\ref{eq19}) and \ (\ref{eq21}) 
self-consistently. This is a mean field approach within the first order 
perturbation theory. The chemical potential cannot exceed the energy 
$E_{\bf i}$ at any temperature. At the transition temperature, the chemical 
potential is equal to the ground state energy. Therefore, the transition 
temperature has to be determined by the chemical potential. We note that the 
second term of Eq.\ (\ref{eq19}) shifts the transition temperature. If we 
choose $U_{{\bf i,j}}$ as a constant i.e. $U$, then it only causes the constant 
shift for all $E_{\bf i}$, but the transition temperature does not change. 
Therefore, the second term of Eq.\ (\ref{eq19}) plays an important role to 
understand the interacting Bose gas. We also note that, when we put $g=0$, 
Eq.\ (\ref{eq23}) reproduces the total number of the noninteracting Bose 
gas confined in the spherical harmonic potential, because there are no 
constraints in between the chemical potential $\mu$ and the energy $E_{\bf i}$. 
Therefore, when we consider the interaction effect, we can also investigate the 
finite size effect at the same time.

\section{Numerical Results}
\label{sec:level3}

We present the numerical results of thermodynamic quantities which appear
in Eqs.\ (\ref{eq23}) and \ (\ref{eq25}). First, we numerically solve the 
chemical potential $\mu$ and the energy $E_{{\bf i}}$ by using 
Eqs.\ (\ref{eq19}) and \ (\ref{eq23}) with the iteration method. Then, we 
obtain $\mu$ and $E_{{\bf i}}$ as a function of $T$ and $N$. If the temperature 
of this system is lowered at fixed $N$, the chemical potential $\mu$ given by 
Eq.\ (\ref{eq23}) increases. It finally reaches the ground state energy $E_{0}$, 
since the chemical potential cannot exceed the ground state energy. Therefore, 
we can define the transition temperature $T_{c}$ as $\mu=E_{0}$ at $T=T_{c}$. 
We note that the transition temperature $T_{c}$ for the noninteracting Bose gas 
confined in the spherical harmonic potential can be given as \cite{ket,hau}
\begin{eqnarray}
T_{c} & = & T_{0}\left(1-\frac{\zeta(2)}{2\zeta(3)^{2/3}}\frac{1}{N^{1/3}}
\right),
\label{eq26}
\end{eqnarray}
where the second term is the finite size effect, and $T_{0}$ denote the 
transition temperature in the thermodynamic limit which can be written as
\begin{eqnarray}
T_{0} & = & \frac{\hbar\omega}{k_{B}}\left( \frac{N}{\zeta(3)}\right)^{1/3}.
\label{eq27}
\end{eqnarray}
Because the interaction effect also shifts the transition temperature $T_{c}$, 
we have to redefine the transition temperature.

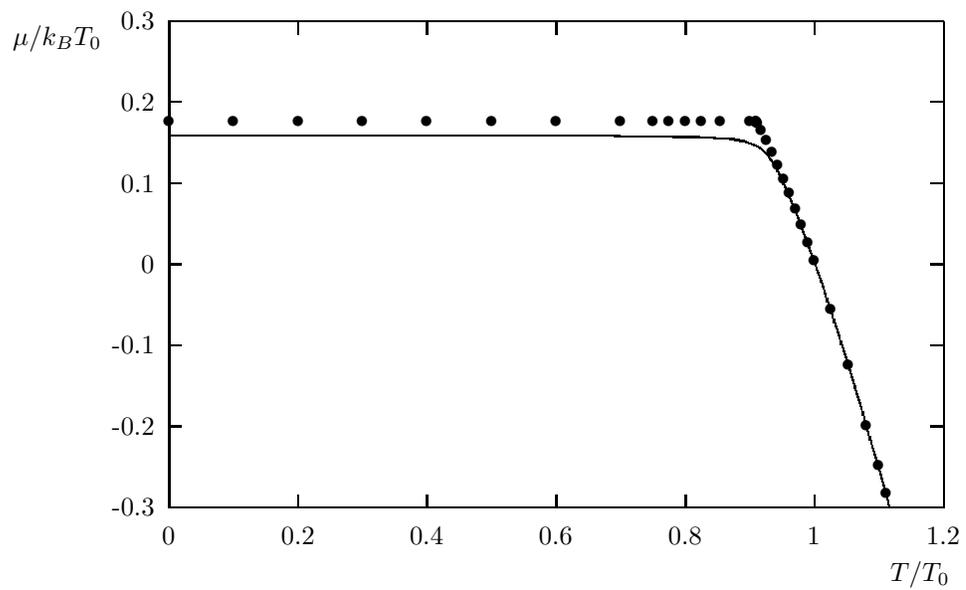
\begin{figure}
\caption{Chemical potential $\mu/k_{B}T_{0}$ as a function of $T/T_{o}$. The 
solid line refers to the noninteracting model. Solid circles show the 
calculated values of the interacting case.}
\label{fig1}
\end{figure}

For $T>T_{0}$, we solve Eqs.\ (\ref{eq19}) and \ (\ref{eq23}) with fixed $N$ 
and given $T$. Here, the total number $N$ and the strength of the interaction 
with the repulsive force are chosen to be $N=1000$ and 
$a/a_{ho}=7\times 10^{-3}$, respectively. In Fig.\ \ref{fig1}, we 
show the calculated values of the chemical potential $\mu /k_{B}T_{0}$ as a 
function of $T/T_{0}$. As can be seen, when $T/T_{0}\sim 0.91$, 
$\mu/k_{B}T_{0}$ approaches to a finite value. This means that the chemical 
potential $\mu$ reaches the ground state energy $E_{0}$. Therefore, the 
transition temperature becomes $T/T_{0}\sim0.91$ (see, Fig.\ \ref{fig3}). 
Comparing this result with the transition temperature $T_{c}/T_{0}\sim 0.93$ 
given by Eq.\ (\ref{eq26}), the interaction effect further shifts the 
transition temperature about $2$ percent toward the lower temperature.

For $T<T_{0}$, the chemical potential $\mu$ is equal to the ground state energy 
$E_{0}$. In this case, since the ground state occupation numbers $N_{0}$ 
diverges, we have to rewrite Eq.\ (\ref{eq19}) as follows
\begin{eqnarray}
E_{{\bf i}} & = & \varepsilon_{{\bf i}}+
\hbar\omega\frac{8\pi a}{a_{ho}}U_{{\bf i},0}N \nonumber \\ 
& & +\hbar\omega\frac{8\pi a}{a_{ho}}\sum_{{\bf j}\neq 0}\langle n_{{\bf j}}
\rangle(U_{{\bf i},{\bf j}}-U_{{\bf i},0}),
\label{eq28}
\end{eqnarray}
where we use the relation 
$N=N_{0}+\sum_{{\bf j}\neq 0}\langle n_{{\bf j}}\rangle$.
Under this condition, we solve Eq.\ (\ref{eq28}) with $\mu/k_{B}T_{0}=0.178$. 
We note that $\mu/k_{B}T_{0}$ is $0.159$ for the noninteracting model at the 
transition temperature, since $\mu$ is $3/2\hbar\omega$ in our approach.

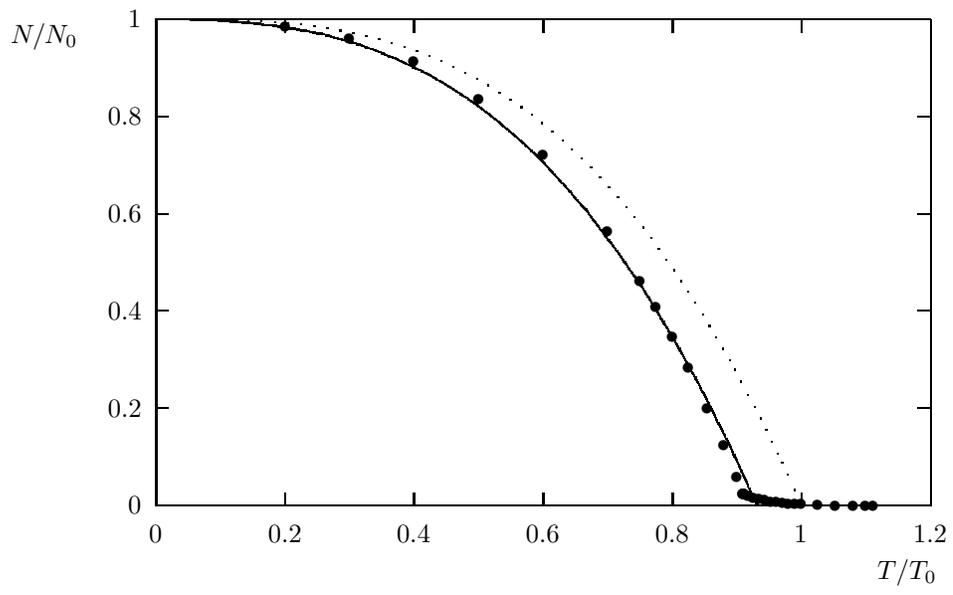
\begin{figure}
\caption{Condensate fraction $N/N_{0}$ as a function of $T/T_{0}$. The solid 
line is drawn with Eq.\ (\protect\ref{eq1}) and the dotted line is the thermodynamic 
limit. Solid circles are the same as in Fig.\ \protect\ref{fig1}.}
\label{fig2}
\end{figure}

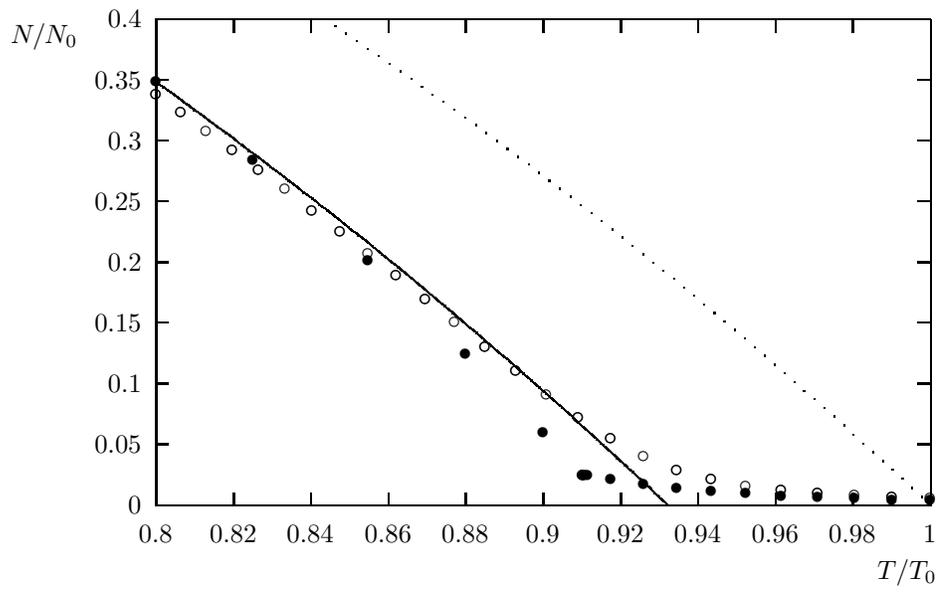
\begin{figure}
\caption{The same as Fig.\ \protect\ref{fig2} just below the transition temperature. 
Open circles are numerical results of the noninteracting case.}
\label{fig3}
\end{figure}

In Fig.\ \ref{fig2}, we show the condensation fraction as a function of 
$T/T_{0}$. The dotted line refers to the condensation fraction of the 
noninteracting Bose gas in the thermodynamic limit, and the solid line 
represents the condensation fraction including the finite size effect 
(see, Eq.\ (\ref{eq1})). Finally, the condensation fraction with the 
interaction and the finite size effects is shown as solid circles. These 
correspond to each other at $T>T_{0}$. This means that we can ignore the 
atom-atom interaction because the number density is low at $T>T_{0}$. On the 
other hand, at near the transition temperature, atoms must occupy the ground 
state energy $E_{0}$. Since the number density becomes high at this 
temperature, the atom-atom interaction becomes strong, and shifts the 
transition temperature. Although the finite size effect can also shift the 
transition temperature, it disappears at large $N$ limit. Therefore, the 
interaction effect is more important than the finite size effect.

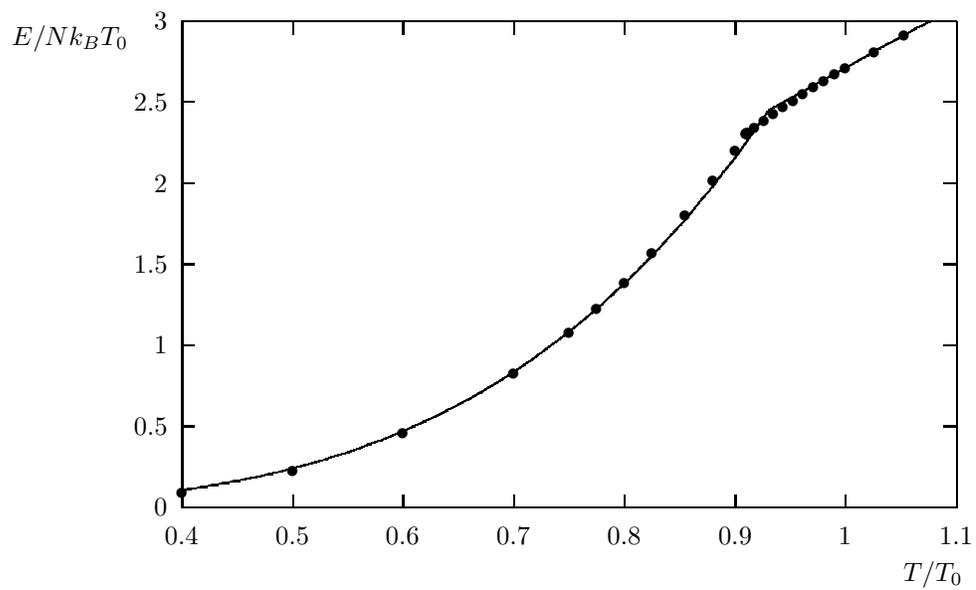
\begin{figure}
\caption{Internal energy $E/Nk_{B}T_{0}$ as a function of $T/T_{0}$. The solid 
line is the noninteracting case. Solid circles are the same as in 
Fig.\ \protect\ref{fig1}.}
\label{fig4}
\end{figure}

In Fig.\ \ref{fig4}, we show the results for the internal energy 
$E/Nk_{B}T_{0}$ as a function of $T/T_{0}$. It reveals that the internal energy 
curve changes slope near the transition temperature. We note that the specific 
heat can be calculated by differentiating the internal energy with respect to 
the temperature $T$. Therefore, this result means that the specific heat is 
discontinuous at the transition temperature. As can be seen from the result of 
Ensher et al. \cite{ens}, the derivative of the internal energy is 
discontinuous at the transition temperature. This behavior of our calculated 
internal energy is in a good agreement with the experimental result.

Finally, we compare our results with the semiclassical results. In the 
semiclassical approximation, the dimensionless parameter 
$E_{int}/E_{kin}=Na/a_{ho}$ is important to discuss the two-body atomic 
interaction energy $E_{int}$ compared to the kinetic energy $E_{kin}$. For the 
Thomas-Fermi (TF) limit $(Na/a_{ho}\gg 1)$, the chemical potential $\mu_{0}$ 
can be written in terms of this parameter \cite{bay,edw}
\begin{eqnarray}
\mu_{0} & = & \frac{\hbar\omega}{2}
\left( 15N\frac{a}{a_{ho}}\right)^{2/5}.
\label{eq29}
\end{eqnarray}
In our case, $\mu_{0}/k_{B}T_{0}$ is 0.342. This is about twice as large as the 
$\mu/k_{B}T_{0}=0.178$. Since we take the parameter $Na/a_{ho}\sim 1$, it does 
not achieve the TF limit. Therefore, the kinetic term plays an important role in 
this value. Further, the interaction term $U_{\bf ij}$ in Eq.\ (\ref{eq19}) 
suppresses the total number $N$ of $Na/a_{ho}$. The shift of the transition 
temperature $\delta T_{0}/T_{0}$ in the semiclassical approximation for the 
large $N$ limit can be given as \cite{giob} 
\begin{eqnarray}
\frac{\delta T_{0}}{T_{0}} & = & -1.3\frac{a}{a_{ho}}N^{1/6}.
\label{eq30}
\end{eqnarray}
In our case, the shift of the transition temperature ${\delta T_{0}}/{T_{0}}$ 
is about 3 percent. Because of the interaction term $U_{\bf ij}$, the 
transition temperature is also suppressed by its term in our model.

\section{Conclusions}
\label{sec:level4}

We have presented the thermodynamic potential of the interacting Bose gas 
confined in the spherical harmonic potential within a mean field approach. In 
this approach, we minimize the interacting Hamiltonian $\hat{H}'$ under the 
condition that the mean occupation numbers $\nu_{\bf i}$ should be the Bose 
distribution function. Then, we can derive the thermodynamic potential in a self-consistent fashion. With this thermodynamic potential, we obtain the 
thermodynamic quantities which have the same form as that of the ideal Bose 
gas. Further, we carry out the numerical calculations of the chemical 
potential, the condensate fraction and the internal energy. In particular, 
because the energy $E_{\bf i}$ also varies as a function of $T$ and $N$ in our 
approach, we have to solve $E_{\bf i}$ and $\mu$ with the conditions that $\mu$ 
cannot exceed $E_{\bf i}$ at any temperature. Further, since we take the 
parameter $Na/a_{ho}\sim 1$ which corresponds neither to the TF limit nor to 
the noninteracting limit for the semiclassical point of view, we cannot expect 
$\mu$ from the semiclassical approximation. Then, we find that the interaction 
effect shifts the transition temperature $T_{0}$ given by $\mu=E_{0}$ at 
$T=T_{c}$ about 2 percent toward the lower temperature compared with the 
noninteracting model including the finite size effect.

\acknowledgments

I thank T. Fujita for careful reading.

\section*{Figure Captions}

\widetext


\begin{references}

\bibitem{and} M. H. Anderon, J. R. Ensher, M. R. Matthews, C.E. Wieman, 
and E. A. Cornell, Science {\bf 269}, 198 (1995).

\bibitem{bra} C. C. Bradley, C. A. Sackett, J. J. Tollett, and R. G. Hulet, 
Phys. Rev. Lett. {\bf 75}, 1687 (1995).

\bibitem{dav} K. B. Davis, M.-O. Mewes, M. R. Andrews, N. J. van Druten, 
D. S. Durfee, D. M. Kurn, and W. Ketterle, \\
Phys. Rev. Lett. {\bf 75}, 3969 (1995).

\bibitem{mew} M. -O. Mewes, M. R. Andrews, N. J. van Druten, D. M. Kurn, 
D. S. Durfee, and W. Ketterle, Phys. Rev. Lett. {\bf 77}, 416 (1996).

\bibitem{ens} J. R. Ensher, D. S. Jin, M. R. Matthews, C. E. Wieman, 
and E. A. Cornell, Phys. Rev. Lett. {\bf 77}, 4984 (1996).

\bibitem{ket} W. Ketterle and N. J. van Druten, 
Phys. Rev. A {\bf 54}, 656 (1996).

\bibitem{hau} H. Haugerud, T. Haugset, and F. Ravndal, 
Phys. Lett. A {\bf 225}, 18 (1997).

\bibitem{gro} E. P. Gross, Nuovo Cimento {\bf 20}, 454 (1961), 
J. Math. Phys. {\bf 4}, 195 (1963).

\bibitem{pit} L. P. Pitaevskii, Sov. Phys. JETP {\bf 13}, 451 (1961).

\bibitem{dal} F. Dalfovo, S. Giorgini, L. P. Pitaevskii, and S. Stringari, 
Rev. Mod. Phys. {\bf 71}, 463 (1999).

\bibitem{gri} A. Griffin, Phys. Rev. {\bf B53}, 9341 (1996).

\bibitem{gioa} S. Giorgini, L. P. Pitaevskii, and S. Stringari, J. Low. 
Temp. Phys. {\bf 109}, 309 (1997).

\bibitem{jin} D. S. Jin, M. R. Matthews, J. R. Ensher, C. E. Wieman, 
and E. A. Cornell, Phys. Rev. Lett. {\bf 78}, 764 (1997).

\bibitem{giob} S. Giorgini, L. P. Pitaevskii, and S. Stringari, Phys. Rev. 
A {\bf 54}, 4633, (1996).

\bibitem{fet} A. L. Fetter, Ann. Phys. (N.Y.), {\bf 70}, 67 (1972).

\bibitem{lan} L. D. Landau and E. M. Lifshitz, Statistical Physics, 3rd ed.
(Pergamon Press, 1980).

\bibitem{bay} G. Baym and C. Pethick, Phys. Rev. Lett. {\bf 76}, 6 (1996).

\bibitem{edw} M. Edwards and K. Burnett, Phys. Rev. {\bf A51}, 1382 (1995).

\end{references}
\end{document}